\documentstyle[12pt,leqno]{article}

 \newcommand{\beq}{\begin{equation}}                       
 \newcommand{\eeq}{\end{equation}}                         
 \newcounter{nt}[section]                                  
 \newcounter{nl}[section]                                  
 \date{}  

\begin{document}

\vspace*{14mm}
\begin{center}
{ \Large \bf Existence and uniqueness results for  a class of  non linear models}
\end{center}

\vspace{2mm}
\begin {center}
{\normalsize Nota di M. De Angelis,  A. Maio, E. Mazziotti}
\end{center}

\begin {center}
 {\small
  Facolt\`a di Ingegneria, Dip. Mat.e Appl., via Claudio 21, 80125, Napoli, Italy.}
\end {center}
\begin{center}
\vspace{-3mm}
 {\small e-mail modeange@unina.it}
\end{center}

\vspace{5mm}

\vspace{8mm}\noindent
{ \small \bf Abstract -}
{\small  The qualitative analysis of the initial value problem  $ \, {\cal P}\,$   related to a {\em non linear} third order parabolic  equation typical of diffusive models is discussed. Some basic properties of the the fundamental solution of a related linear operator are determined and are applied to an equivalent integro differential formulation  of the problem. By the fixed point theorem, existence and uniqueness results are obtained.}


\vspace{4mm}\noindent
{\small \it \bf Key words:}
{\small   Diffusion models, Partial and integro differential equations, Laplace transform }

\vspace{6mm}
 \section{  Introduction}
 \setcounter{equation}{0}

\hspace{5.1mm}

\vspace{3mm}

Several mathematical models in applied sciences - as Biochemistry, Epidemiology, Population dynamics - deal with higher order parabolic equations typical of diffusive phenomena. The results of experimental analysis  show that the diffusion coefficients are generally {\it small} and this requires too much time in the transmission of signals. As consequence it necessary to correct the diffusive models by means of reaction-diffusion systems. So, one obtains non linear hyperbolic equations perturbed by viscous terms of higher order. 


A typical example is given by equations as

\vspace{3mm}

\beq                                                     \label{11}
   L u \equiv \ (\,\varepsilon \partial _t + c^2  \,)\, u_{xx} - (\,\partial_{t}\,+\,a\,)\, u_t  = \ F(\,x,t,u,u_x\,),
\eeq

\vspace{3mm}\noindent
which are object of an  extensive literature \cite {ar}- \cite{hi}.

When $\, F \, $ is linear, numerous contributions are  already known  \cite {n} - \cite{dfm}. In particular the fundamental solution  $\,K \, $ of the operator $\, L\, $ has been obtained in \cite {dm} where some basic properties has been discussed too.

 In this paper, the non linear case of (\ref{11}) will be considered and existence and uniqueness results for the initial value problem   $ {\cal P}$ in all of the space will be obtained.

For this, the equivalence between the problem  $ {\cal P}$ and an integro - differential equation with kernel $ \, K\, $ is deduced and the fixed point theorem is applied by means  of appropriate estimates for $ \, K .\, $ 

This analysis can be also applied to obtain the  dependence continuously upon the data, stability properties and boundary - layer estimates. More, the study of other boundary value problems (semi infinite or finite media) with Dirichlet or Neumann conditions could be achieved too.

 \vspace{5.1mm}

\section {Fundamental solution and its properties}
 \setcounter{equation}{0}

\vspace{10mm}

If $\   D =\{(x,t) :  x \in  \  \Re\ , \ 0 < t \leq T \ \,\},$ let $r = |x|/\varepsilon, \,\, b= c^2/\varepsilon. \,$ As one can verify, the fundamental solution of the $L-$ operator is simbolically defined by 

\vspace{3mm}

\begin{equation}                           \label{21}
\hat K (r,s) = \
\frac{e ^{ \,-\, |x| \, \sqrt {s(s+a ) / (\varepsilon s + c^2) }}}{ 2 \, \sqrt{ s (s+a) (\varepsilon
s +c^2)}}\,\,.
\end{equation}

\vspace{3mm} \noindent
If one puts
 \vspace{3mm}

\beq      \label{22}
  \,\hat G(r,s)\,\,=\
\frac{e ^{ \,-\, r \, \sqrt {s(s+a ) / ( s + b) }}}{ 2 \,  \sqrt \varepsilon  \, \sqrt{  (s+a) (s +b)}} \,\,\,,
\eeq

\vspace{3mm}
\noindent obviously one has:

\vspace{3mm}
\begin{equation}                           \label{23}
\hat K \,(r,s) = \
\frac{1}{\sqrt s}\, \, \hat G ( r,s) \,\, \Rightarrow \,\, K  \,(r,t) \, = \, \int^t _0 \,  G( r,\tau) \, \,\frac{d\tau}{\sqrt{\pi (t-\tau)}}
\end{equation}

\vspace{3mm}\noindent where $\,G ( r,t)\,$ is the inverse tranform of $\, \hat G ( r,s) \,$ given by (\ref{22}).

 If $ I_n (z) \,$  denotes the  modified Bessel function of  the first kind, the explicit expression of $\,G ( r,t)\,$  is given by :

\vspace{3mm}

\beq                 \label{24}
G(r,t) = \, \frac{r}{4 \, \sqrt{\pi \, \varepsilon}} \,\, \int^t_0 \,\, \frac{e^{- \frac{r^2}{4v}}}{  v\,\sqrt{v}}  \, \,\,e^{-\,b (t-v)}\,\, I_0 (r \sqrt{(b-a)(t-v)/v}\,\,\,dv,
\eeq

\vspace{3mm} \noindent and the following theorem holds:

\vspace{5mm}

{\bf Theorem 2.1}- {\em For all  }$\,r>0,\,$  {\em the Laplace integral}  $\,{\cal L  }_t\,\,G(r,t)\,$ {\em converges absolutely
in the half-plane} $ \Re e  \,s > \,max(\,-a,\,-b\,),\,$ {\em and one has:}

 \beq      \label{25}
\,{\cal L  }_t \, G(r,t)\,\,=  \,\hat G(r,s)\,\,=\
\frac{e ^{ \,-\, r \, \sqrt {s(s+a ) / ( s + b) }}}{ 2 \,  \sqrt \varepsilon  \, \sqrt{  (s+a) (s +b)}} \,\,.
\eeq 
\hbox{}\hfill\rule{1.85mm}{2.82mm}

\vspace{3mm}\noindent
According to the results extablished in \cite{dm}, for the function $\,K\, $ the following properties hold :

\vspace{5mm}

{\bf Theorem 2.2}- {\em The function  }$\,K\, $  {\em
defined by }(\ref{23}), (\ref{24}) {\em is   a  } $ \, C^\infty (D) \,$ {\em solution of the equation } $\, L\,u\,=\,0.\,$ {\em When }$ a<b=c^2/\varepsilon, \,\,\, K \,$ {\em is  never negative  in } $ D \,$ {\em and more it results}:

\vspace{3mm}
\begin{equation}                  \label {26} 
  0\,\leq\, \, \,\int_ {-\infty} ^\infty \,\,\, K \,\,(\,|x-\xi|,t\,) \,\,d\xi \, = \,(\,\frac{1}{a}\ )\,\,(1-e^{-\,at})\,\leq \,1/a,
\eeq
\vspace{3mm}

\begin{equation}                  \label {27} 
  \lim_{r \to 0}    \,\, \partial_ r\, K (r,t) \, =\, - \,\,\frac{e^{\,- \,b\,t\,}}{2 \varepsilon}.
\eeq 
\hbox{}\hfill\rule{1.85mm}{2.82mm}

\vspace{3mm}\noindent Moreover, one has:

\vspace{5mm}

{\bf Corollary 1.2  -} {\em By means of the Laplace transforms, it's easy to verify the following formulae:}

\vspace{3mm}
\begin{equation}                  \label {28} 
  \int_\Re \,\,\, (\, \partial _t \, + a \, )  K \,\,(x,t) \,\,dx \, = \,1,
\eeq
\vspace{3mm}

\begin{equation}                  \label {29} 
\int_\Re \,\,\, (\, \partial _t \, + b \, )  K \,\,(x,t) \,\,dx \, = \, e^{ \, - \, a \, t\, } \, + \, b\, ( 1 \, - \,\,e^{-\,at})\,/ a \,\,,
\eeq

\vspace{3mm}
\begin{equation}                  \label {210} 
  \int_\Re \,\,\, (\, \partial _t \, + a \, -\,\varepsilon \, \partial_{xx}^2 \, )  K \,\,(x,t) \,\,dx \, = \,1\, - \,  e^{ \, - \, b \, t\, } \,.
\eeq 
\hbox{}\hfill\rule{1.85mm}{2.82mm}

\vspace{10mm}

\section{Initial  value problem and explicit solution in the linear case}       
\setcounter{equation}{0}

\vspace{5mm}
Let consider  now the  initial value problem $ {\cal P}$ in $\, D \,$: 
\vspace{3mm}

\beq                                                     \label{31}
  \left \{
   \begin{array}{ll}
    & L u \equiv \ 
    (\varepsilon \partial _t + c^2  \,)\, u_{xx} - (\partial_{t}+a) u_t  = \ F(x,t,u,u_x),\ \  \
       (x,t)\in D,\vspace{2mm}\\
\\
   & u (x,0)= f_0(x), \  \    u_t (x,0)= f_1(x),
   \ \ \ \ \ \ \ \ \ \ \ \ \ \ \ \ \  \ \ \ \ \ \ \ \ \ \   x \in \Re ,\vspace{2mm}  \\
       \end{array}
  \right.
 \eeq

\vspace{3mm}
\noindent
where  the initial data $\, f_0 \,$ and  $\, f_1 \,$are arbitrary specified functions.

\vspace{3mm}
If  $\, g(x)\,$ is a continuous function on $\Re$, consider the convolution:

\vspace{3mm}
\begin{equation}                  \label {32} 
  u_g \, =\, \int_\Re \,\,\,  g(\xi)\,  K \,\,(x-\xi,t) \,\,d \xi \, = \, K * g.
\eeq


\vspace{3mm}\noindent
Then, owing to the properties of $\, K(x,t),\,$ it's possible to prove the following theorem:

\vspace{5mm}

{\bf Theorem 3.1}- {\em  If} $\, \alpha , \,\beta,\, \gamma \,$ {\em   are three positive constants such that }

\vspace{3mm}
\beq        \label{34}
| g(x) | <\,\, \alpha \,\,\, e^{ \,\, \beta \, |x| ^{\,\gamma\,+1\,}\,} \,\,\,\,\,\,\, \ \ \ \\ \ \\ \ \ \ \ \ \ \ \ \ \   \ \ 0\leq  \gamma  < \, 1 \, ,
\eeq

\vspace{3mm}\noindent {\em  then the function } $\,u_g\, $   {\em is   a  } $ \, C^\infty (\,D\,) \,$ {\em solution of the equation } $\, L\,u\,=\,0,\,$ 
{\em such that }

\vspace{3mm}
\beq              \label{35}
\lim _{t \to 0}\,\, u_g(x,t) \, = \,0,    \,\,\,\,\,\,\, \ \ \ \ \lim _{t \to 0}\,\, \partial_t \,u_g(x,t) \, = \,g(x)  \,\,\,,                        
\eeq 

 \vspace{3mm}\noindent {\em uniformly for all }$\, x\, $ {\em in a compact subset of }   $\, -\infty \, <\, x\,< \, \infty \,$. 
\hbox{}\hfill\rule{1.85mm}{2.82mm}

 \vspace{5mm}
Observe now that the convolution $ \, K*g\,$ is infinitely differentiable in $\, D\, $ and,  according to the results of  section 2, one has:

\vspace{3mm}
\begin{equation}                  \label {36} 
( \partial _t \, + a  -  \varepsilon  \partial_{xx}^2 \, ) u_g \,= e ^ {\, -\, b\, t\, }\, g(x) \, + \, \int_\Re \,\,\,g(\xi)\,\, (\, \partial _t \, + a \, -\,\varepsilon \, \partial_{xx}^2 \, )  K(x-\xi,t)  \,\,  d \xi \,
\eeq

\vspace{3mm}\noindent and  (\ref{210}) implies:

\vspace{3mm}
\begin{equation}                  \label {37} 
\lim _{t \to 0 } \, \,\, \int_\Re \,\,\,  g(\xi) \,\, (\, \partial _t \, + a \, -\,\varepsilon \, \partial_{xx}^2 \, ) \,\, K(x-\xi,t) \,\,  d \xi \, =\,\,0.
\eeq

\vspace{3mm}\noindent Consequently, one has :

\vspace{5mm}

{\bf Theorem 3.2 -} {\em  If }$ \, g(x)\,$  {\em verifies the hypotheses of Theorem} 3.1 , {\em then  the function }

\vspace{3mm}
\beq        \label{38}
u_g^* \, =(\, \partial _t \, + a \, -\,\varepsilon \, \partial_{xx}^2 \, )\,\, u_g
\eeq

\vspace{3mm}\noindent {\em represents a smooth solution of  the equation } $\, L\,u\,=\,0,\,$ 
{\em such that }

\vspace{3mm}
\beq              \label{39}
\lim _{t \to 0}\,\, u^*_g(x,t) \, = \,g(x),    \,\,\,\,\,\,\, \ \ \ \ \lim _{t \to 0}\,\, \partial_t \,u^*_g(x,t) \, = \,0                         
\eeq 

 \vspace{3mm}\noindent {\em uniformly for all }$\, x\, $ {\em in a compact subset of }   $\, -\infty \, <\, x\,< \, \infty \,$. 
\hbox{}\hfill\rule{1.85mm}{2.82mm}

\vspace{5mm} 

{\bf Remark 3.1 -} {  When $\, F \equiv 0 \,, \, $ the results of Theorems} 3.1 - 3.2 allow to assert that the function

\vspace{3mm}
\beq        \label{310}
u(x,t) \,= u_{f_1}  \,+\, (\, \partial _t \, + a \, -\,\varepsilon \, \partial_{xx}^2 \, )\,\, u_{f_0}
\eeq

\vspace{3mm}\noindent { represents a smooth solution of  the  homogeneous initial - value problem } (\ref{31}).
\hbox{}\hfill\rule{1.85mm}{2.82mm} 

\vspace{5mm} 
  Consider now the case that $\,F\,$ is not vanishing but  it is a {\em linear known } function   $\,F = f(x,t) \,$ and let $\,u_f \,$ the {\em  volume potential}

\vspace{3mm}
\begin{equation}                  \label {311} 
  u_f \, = \, \int _0^t \, K * f \, \ d \tau \,= \int _0^t \, \ d \tau \, \int_\Re \,\,\,  f(\xi, \tau )\, \, K \,\,(x-\xi,t-\tau) \,\,d \xi \,\,\ . 
\end {equation}

\vspace{3mm}\noindent  It's possible to prove the following

\vspace{7mm}

{\bf Theorem 3.3 -} {\em When the source term } $ f \,(x, t\,) \,$ {\em is  continuous and  everywhere bounded function in the set } $  -  \infty <\xi <\infty ,\,\, 0<\tau <T,  \,$ 
{\em then } $\,u_f \,$ {\em represents a smooth solution of the   problem}  $ L \,u_f = -\, f(x,t)\,\,$ {\em with vanishing initial conditions.}

\vspace{3mm}\noindent Proof - If $\,\,\,||\,f\,||_T \,= \sup _{D}\, | \,f \,(\,x, t\,) \,|, \,\,$ by (\ref{26}) one has:

\vspace{3mm}
\begin{equation}                  \label {312}
 |\, u_f\,(x,t)\, |\leq \, a^{-1} \,\, t\,\, ||\,f\,||_T \, \, \Rightarrow \,\,  \lim_ {t \to 0 } \, u_f \\  = \,0 \,\,. 
\eeq

\vspace{3mm}\noindent  Further it results:

\vspace{3mm}
\begin{equation}                  \label {313}
 |\,K * f\,\, |\,\,\ \leq \, a^{-1} \,\, [\, 1- e^{-\,a(t-\tau)}\,]\,\, ||\,f\,||_T 
\eeq

\vspace{3mm}\noindent and so 

\begin{equation}                  \label {314}
 \lim _{\tau  \to t } \, \int_\Re \,  K(x-\xi,t-\tau) \,\,f( \xi,\tau,)  d \xi \, =\,\,0 \,\, \,\, \Rightarrow   \lim _{\tau  \to 0 } \, \partial _t \, u_f \,=\,0
\eeq

\vspace{3mm}\noindent  More, by means of estimates similar  to (\ref{37}), one deduces that:

\vspace{3mm}
\beq         \label{315}
\lim _{\tau \to t } \, \,\, \int_\Re \,\,\,  f(\xi, \tau) \,\, (\, \partial _t \, + a \, -\,\varepsilon \, \partial_{xx}^2 \, ) \,\, K(x-\xi,t-\tau) \,\,  d \xi \, =\,\,0,
\eeq

\vspace{3mm}\noindent  and this, together with the properties of $\, K\,$ (theorem 2.2), implies $ L \,u_f = -\, f(x,t).\,\,$ 

\vspace{3mm}\noindent Further, theorems 3.1 - 3.2 imply that $\, u_f\, (x,0)\,=\,0,  \ \,\,\partial_t \,u_f (x,0)\,=\,0.$
\hbox{}\hfill\rule{1.85mm}{2.82mm}

\vspace{5mm}
At last, by (\ref{310}) and theorem 3.3, it follows:

\vspace{5mm}

{\bf Theorem 3.4 -} {\em When  } $ \, F\,= \,f \,(x, t\,) \,$ {\em and the data }$ \, f,\,\, \,f_0, \,\,\,f_1 \, \,$ {\em verify the hypotheses specified in theorems } 3.1 - 3.3, {\em then an explicit  smooth solution of the problem }(3.1) {\em is given by }

\vspace{3mm}
\beq       \label {316}
u(x,t) \,=  \,\, - u_f \, + \, u_{f_1}  \,+\, (\, \partial _t \, + a \, -\,\varepsilon \, \partial_{xx}^2 \, )\,\, u_{f_0}
\eeq
\hbox{}\hfill\rule{1.85mm}{2.82mm}

\vspace{5.1mm}

\section{  Integro differential formulation of the non linear problem}       
\setcounter{equation}{0}

\vspace{5mm}

 Consider now the non linear case of the problem  (\ref{31}), where $\, F\,=\,F\,(\,x,t,u,u_x\,)\,$  is defined on the set 

\vspace{3mm}

\begin {center}
$\ \Omega  \,= \{ \, (\,x,t,u,p\,): ( \, x,t\,) \in D, \,\,\,\,\,-\infty <u<\infty  \,\,;\,\,\,\,-\infty <p<\infty \,\, \}, $
\end{center}

 \vspace{3mm}\noindent with $\   D =\{(x,t) :  x \in  \  \Re\ , \ 0 < t \leq T \ \,\}.$

\vspace{7mm} 
From now on we shall assume for  $\, F= \,F (x,\,t, \,u(\,x,t\,), \,\, u_x(\ x,t\,)\,\, ) \,\,\,$  the following  {\bf Hypotheses H}:

\vspace{5mm}\noindent - The function $\, F (x,t,u,p)\,\,$ is defined and continuous on $\ \Omega  \,$ and  it's bounded for all $ \,u\, $ and  $\, p .\,$

 \vspace{3mm}\noindent -  For each $  \,k\,>\,0\,$ and for $\, |u|,\,|p| <\,k,\,$  the function  $\, F \,\,$ is Lipschitz continuous in    $ \, \,x\,  $ and  $ \,t\,$ for each compact subset of $\, D.\,$

\vspace{3mm}\noindent - There exists a constant $ \, \beta _F \, $ such that:
\vspace{3mm}\begin {center}
 $\, |F (x,t,u_1,p_1)\,-\,F (x,t,u_2,p_2)|\, \leq \,\beta _F \,\, \{\, | u_1-u_2\,| \, +\,| p_1-p_2\,| \, \} \ $ 
\end{center}
\vspace{3mm}
\noindent holds for all  $\,( u_i,\,p_i) \,\, i=1,2. \,$

\vspace{7mm}
Let  $\,u\,$  be a smooth solution of the problem  (\ref {31}) and,  referring to the  function   $ \,\,\, K(x-\xi , \,\, t- \tau \,) \,\,\ u (\xi, \tau), \,\,\,\ $  consider the operators  $ \,\,\, {\cal M}(x, \, t, \, \xi, \, \tau,)\,, \ $  $ \,{\cal N}(x, \, t, \, \xi, \, \tau,)\,\,\,$ defined by:

\vspace{3mm}

\beq     \label{41}
\begin{array}{ll}
                                                     
 &{\cal M}(x, \, t, \, \xi, \, \tau,) =  \varepsilon  \, K \, u _{\xi \xi} 
 \,-  K \, u _{\tau} + u \,  K_{\tau}\, -\, a K \, u  \\

\\
&{\cal N}(x, \, t, \, \xi, \, \tau,) = u \, ( \,\varepsilon  \partial\tau  \, - c^2 \, ) K_\xi \, - u_ \xi  ( \, \varepsilon  \partial_\tau  \, - c^2 \, ) K \,  
\end {array}
\eeq

\vspace{3mm}
Further, let $ \,\,\widetilde{L}$ \ be  the adjoint operator of $L$:

 \vspace{3mm}

\beq                                                     \label{42}
 \widetilde { L} \, K  = - \varepsilon  \, K_{\xi \xi \tau} 
+ c^2  \, K_{\xi \xi} - K_{\tau \tau}\, + \, a K_\tau 
\eeq

\vspace{3mm}\noindent in order to have:

\vspace{3mm}
\beq                                                     \label{43}
 K \, L \,u \, -\, u\, \widetilde { L} \, K  \,= \, K \, F.
\eeq
\vspace{3mm}\noindent As consequence, one has

\vspace{5mm}
\beq                \label{44}
\partial_ \tau {\cal M} \, + \partial _\xi \, {\cal N} \, = K F .
\eeq

\vspace{3mm}
 If $\,u_F\,$ denotes the non linear potential volume

\vspace{3mm} 
\beq             \label{45}
 u_{F}  \,\, = \,  \int_0^t d\tau \,\int_\Re   K  (\xi-x, \,t -\tau \, )\,\,\, F(\xi,\tau, \, u(\xi,\tau ), \,u_\xi ( \xi ,\tau )\,\,)\,\ d\xi\,, 
\eeq

\vspace{3mm}\noindent  it suffices to integrate (\ref{44}) on $\, D\,$  in order to have

\vspace{3mm}
\beq                                   \label{46}
   \int_{-\infty}^{\infty} \,\, [\,\,{\cal M}\,(\,x, \,t, \, \xi, \, t\,) \, - \,{\cal M}(\,x, \, t, \, \xi,  \,0\,)\, \,] \,\,d\xi  \,  =  u_F
\eeq

\vspace{3mm}\noindent where the term with $\partial_\xi \,{\cal N}$ doesn't contribute because $ \,\,\lim _{|\xi | \to \infty} \ \,{\cal N} =0\,$.

\vspace{5mm}
Let  assume that  $\, f_0 \in C^2\,(\Re),\,\,f_1\, \in C^1\,(\Re), \,$ and more $\,f_0,  f'_0, \, \, f''_0, \,\, f_1 \,\,$ let bounded on $ \Re.\,\,\ $  Then, by means of (\ref{35}) - (\ref{37}) and estimates similar to (\ref{314}),(\ref{315}), by (\ref{46}) one obtains the following representation

\vspace{3mm} 
\beq             \label{47}
u(x,t) = \,\, -u_{F}\,\, +  \int_\Re \, \,f_1(\xi)\,\, K  (\,\xi-x,\,t \,)\,\,\,d\xi \,\, +  
\eeq
\\
\[ +\,\, ( \,\partial _t \, +\,a - \,\varepsilon \,\,\partial_ {xx}\,)\, \int_\Re  \,\,f_0(\xi)\,\, K  (\,\xi-x,\,t\, )\,\,\, d\xi  \,\, \]

\vspace{5mm}
Viceversa, suppose that the integral differential equations (\ref{47}) has a solution $ u= u(x,t)$ such that $ u $ and $ u_x$  are continuous and bounded  for $ x \in \Re $ and $0 < t\leq T$. Then the terms depending on the initial data $ f_0,\,\, f_1 \, $ are differentiable  for $\,t>0\,$ with bounded derivatives  for $\,t>0\,$. The properties of  the fundamental solution $ \, K\, $ and the hypotheses H allow to prove that also the  volume potential $\,u_{F}\,$  and its derivatives are  Lipschitz continuous. Then  the function $ \,u\,$  characterized  by (\ref{47}) verifies the problem (\ref {31}) everywhere in $D$. 

\vspace{3mm}The complete equivalence between  the problem  ${\cal P}$ and the integro differential equation (\ref{47}) confirms the existence and uniqueness of  the solution of  (\ref {31})  as soon as when this will be demostrate for (\ref{47}).

 \vspace{5.1mm}

\section {Existence and uniqueness}
 \setcounter{equation}{0}

\hspace{5.1mm}

\vspace{3mm}
Consider now a time - interval $[0,\eta]$, with $\eta <T$. At first we will show that the solution of the integral differential equation (\ref{47}) exists and is unique for $ t\, \in \,[0,\eta]$ and then we will extend this for any finite $T$.

For $\eta >0$, let  

\beq     \label{51}
   D_\eta =\{(x,t) :  x \in  \  \Re, \ \  t \in [0,\eta] \ \}
\eeq

\vspace{3mm}\noindent and let   ${\cal B}_\eta$
  the space

\vspace{3mm}
\beq     \label{52}
   {\cal B}_\eta = \{v(x,t) :  v,\,\, v_x \in \\ \, C(D_\eta)\}
\eeq

\vspace{3mm}\noindent
equipped with the norm

\vspace{3mm}
\beq     \label{53}
   ||v||_\eta \,=\, \sup_{D_\eta} \, |v(x,t)| \, + \sup_{D_\eta} \, |v_x(x,t)| \, <\infty.
\eeq

\vspace{3mm}\noindent

The set    is a Banach space.
The arguments discussed previously allow to state that the  mapping

\vspace{3mm}
\beq    \label{54}
{\cal F} v(x,t)= \int_\Re f_1(\xi) K  (\xi-x,t )d\xi  +  ( \partial _t \, +a - \varepsilon\partial_ {xx}) \int_\Re  f_0(\xi) K  (\xi-x,t ) d\xi +
\eeq
\[  + \,\,\int_{0}^{t} \, \,d \tau \,\int_{-\infty}^{\infty}  \, K(x-\xi, t-\tau ) \, F(\xi, \tau ,\, v(\xi,\tau), \, v_\xi(\xi, \tau)) \ d \xi\]

\vspace{5mm}\noindent maps $ {\cal B}_\eta $ into $ {\cal B}_\eta .$

\vspace{3mm}Estimate  now the continuity of $ {\cal F} $. For all $t \in [0,\eta]$, owing to (\ref{26}) and hypotheses H , it results

 \vspace{3mm}
\beq    \label{55}
|{\cal F} v_1(x,t)-{\cal F} v_2(x,t)| \, \leq  \beta_F \, \,  a^{-1}\,\, t \,  ||v_1-v_2||_\eta  \leq \,  \beta_F \, \, a^{-1}\,\, \eta \,\, ||v_1-v_2||_\eta  .
\eeq

\vspace{3mm}\noindent
As for the x - derivative, by means of  $ \,\hat K_r (z,t)\,$, one can prove that:

\vspace{3mm}
\beq    \label{56}
|\frac{\partial}{\partial_x}\,{\cal F} v_1(x,t)- \frac{\partial}{\partial_x}\,{\cal F} v_2(x,t) |\, \leq \,\, \beta_F \, \,\, a^{-1} \,\, \frac{1}{\sqrt{\varepsilon (b-a)}}\,\, \eta \, \,\, ||v_1-v_2||_\eta  .  
\eeq

\vspace{3mm}\noindent

As consequence, by (\ref{53}), (\ref{55}),(\ref{56}) it follows 

\vspace{3mm}
\beq    \label{57}
||{\cal F} v_1(x,t)-{\cal F} v_2(x,t)||_\eta \, \leq  \beta_F\,\, ( \frac{1}{a}\, + \frac{1}{\sqrt{\varepsilon (b-a)}}) \, \,  \eta \,\,  ||v_1-v_2||_\eta  .
\eeq

\vspace{3mm}\noindent

If we select $\eta$ such that 

\vspace{3mm}
\beq    \label{58}
 \beta_F\,  \,\, ( \frac{1}{a}\, + \frac{1}{\sqrt{\varepsilon (b-a)}}) \,  <1 ,
\eeq

\vspace{3mm}\noindent
then ${\cal F}$ is a contraction of  $ {\cal B}_\eta $ into  $ {\cal B}_\eta $ and so  $ {\cal F} $  has a unique fixed point $u(x,t) \in   {\cal B}_\eta $ .

\vspace{3mm}In order to show the existence and uniqueness of the solution in all $[0,T]$ we proceede by induction.

\vspace{3mm}\noindent Let assume that the equation (\ref{47}) admits a unique solution $\,u\,$ which is bounded for $0\, \leq \, t \, \leq k\eta\, $ together with £ \,$ u_x\, $ . 

For $ \, k\eta \, \leq \, t \, \leq (k+1)\eta,  \, $\, let consider the space

\vspace{3mm}
\beq     \label{59}
   \widetilde{\cal B}_\eta = \{ \, v(x,t)\, : \, v\,(x,t-k\eta)\, \in \\ {\cal B}_\eta \}
\eeq

\vspace{3mm}\noindent and the mapping 

\vspace{3mm}
\beq    \label{510}
{\cal F}_1 v(x,t)= \int_\Re f_1(\xi) K  (\xi-x,t )d\xi  +  ( \partial _t  +a - \varepsilon \partial_ {xx}) \int_\Re  f_0(\xi) K  (\xi-x,t ) d\xi +
\eeq 
\[+ \int_{0}^{k\eta} \, \,d \tau \,\,\int_{-\infty}^{\infty}  \, K(x-\xi, t-\tau ) \, F(\xi, \tau ,\, u(\xi,\tau), \, u_x(\xi, \tau)) \ d \xi  \,+\]
\[  +\int_{k\eta}^{t} \, \,d \tau \,\,\int_{-\infty}^{\infty}  \, K(x-\xi, t-\tau ) \, F(\xi, \tau ,\, v(\xi,\tau), \, v_x(\xi, \tau)) \ d \xi,\]

\vspace{3mm} \noindent
which maps $ \widetilde{\cal B}_\eta$ into $\widetilde{\cal B}_\eta$. Moreover, also the mapping $\,{\cal F}_1\,$ satisfies estimates like (\ref{57}).

\vspace{3mm}
As consequence, for all $\,t \in [\,0,\,(k+1)\eta\,]$ the mapping $\,{\cal F}_1 \,$is a contraction  and hence admits an unique fixed point $\, u \, \in \widetilde{\cal B}_\eta$.

 We remark that , as (\ref{510}) shows, the functions $ \, u \, $ an $\, u_x \, $ are continuous  also on the junction $ \,t = K\eta. \,$

\vspace{3mm}
In conclusion, the results  in sections 4 and 5 allow to state the following theorem:

\vspace{3mm}
{\bf Theorem 5.1 - }{\em Let the initial data such that }$\, f_0 \in C^2\,(\Re),\,\,f_1\, \in C^1\,(\Re). \,$ Let $\,f_0,  f'_0, \, \, f''_0, \,\, f_1 \,\,$ {\em  bounded on} $\, \Re.\,$ {\em Further, let   }$\, F \, $ {\em verify the hypotheses H . Then, the non linear problem }(3.1)  {\em admits a unique  regular solution in D.  }

\hbox{}\hfill\rule{1.85mm}{2.82mm}  

\hspace{5.1mm}

\end{document}